\documentclass[sigconf]{acmart}

\AtBeginDocument{%
  \providecommand\BibTeX{{%
    \normalfont B\kern-0.5em{\scshape i\kern-0.25em b}\kern-0.8em\TeX}}}

\setcopyright{acmcopyright}
\copyrightyear{2022}
\acmYear{2022}
\setcopyright{acmlicensed}\acmConference[ASE '22]{37th IEEE/ACM International Conference on Automated Software Engineering}{October 10--14, 2022}{Rochester, MI, USA}
\acmBooktitle{37th IEEE/ACM International Conference on Automated Software Engineering (ASE '22), October 10--14, 2022, Rochester, MI, USA}
\acmPrice{15.00}
\acmDOI{10.1145/3551349.3559496}
\acmISBN{978-1-4503-9475-8/22/10}

\newcommand{\fg}[1]{{#1}}

\begin{document}

\title{MV-HAN: A Hybrid Attentive Networks based Multi-View Learning Model for Large-scale Contents Recommendation}




\author{Ge Fan}
\affiliation{%
  \institution{Tencent Inc.}
  \city{Shenzhen}
  \country{China}}
\email{ge.fan@outlook.com}

\author{Chaoyun Zhang}
\affiliation{%
  \institution{Microsoft Research}
  \city{Beijing }
  \country{China}}
  \authornote{This work was done when Chaoyun Zhang worked at Tencent.}
\email{chaoyun.zhang@microsoft.com}

\author{Kai Wang}
\affiliation{%
  \institution{Tencent Inc.}
  \city{Shenzhen}
  \country{China}}
\email{wangjinjie722@gmail.com}

\author{Junyang Chen}
\affiliation{%
  \institution{Shenzhen University}
  \city{Shenzhen }
  \country{China}}
  \authornote{Corresponding author.}
\email{junyangchen@szu.edu.cn}

\renewcommand{\shortauthors}{Ge Fan et al.}

\begin{abstract}

Industrial recommender systems usually employ multi-source data to improve the recommendation quality, while effectively sharing information between different data sources remain a challenge. 
In this paper, we introduce a novel \textbf{M}ulti\textbf{-V}iew Approach with \textbf{H}ybrid \textbf{A}ttentive \textbf{N}etworks (MV-HAN) for contents retrieval at the matching stage of recommender systems. The proposed model enables high-order feature interaction from various input features while effectively transferring knowledge between different types. 
By employing a well-placed parameters sharing strategy, the MV-HAN substantially improves the retrieval performance in sparse types.
The designed MV-HAN inherits the efficiency advantages in the online service from the two-tower model, by mapping users and contents of different types  into the same features space. This enables fast retrieval of similar contents with an approximate nearest neighbor algorithm.
We conduct offline experiments on several industrial datasets, demonstrating that the proposed MV-HAN significantly outperforms baselines on the content retrieval tasks. Importantly, the MV-HAN is deployed in a real-world matching system. Online A/B test results show that the proposed method can significantly improve the quality of recommendations. 

\end{abstract}


\maketitle

\section{Introduction}
\begin{figure}[htp]
\centering
\includegraphics[width=8.5cm]{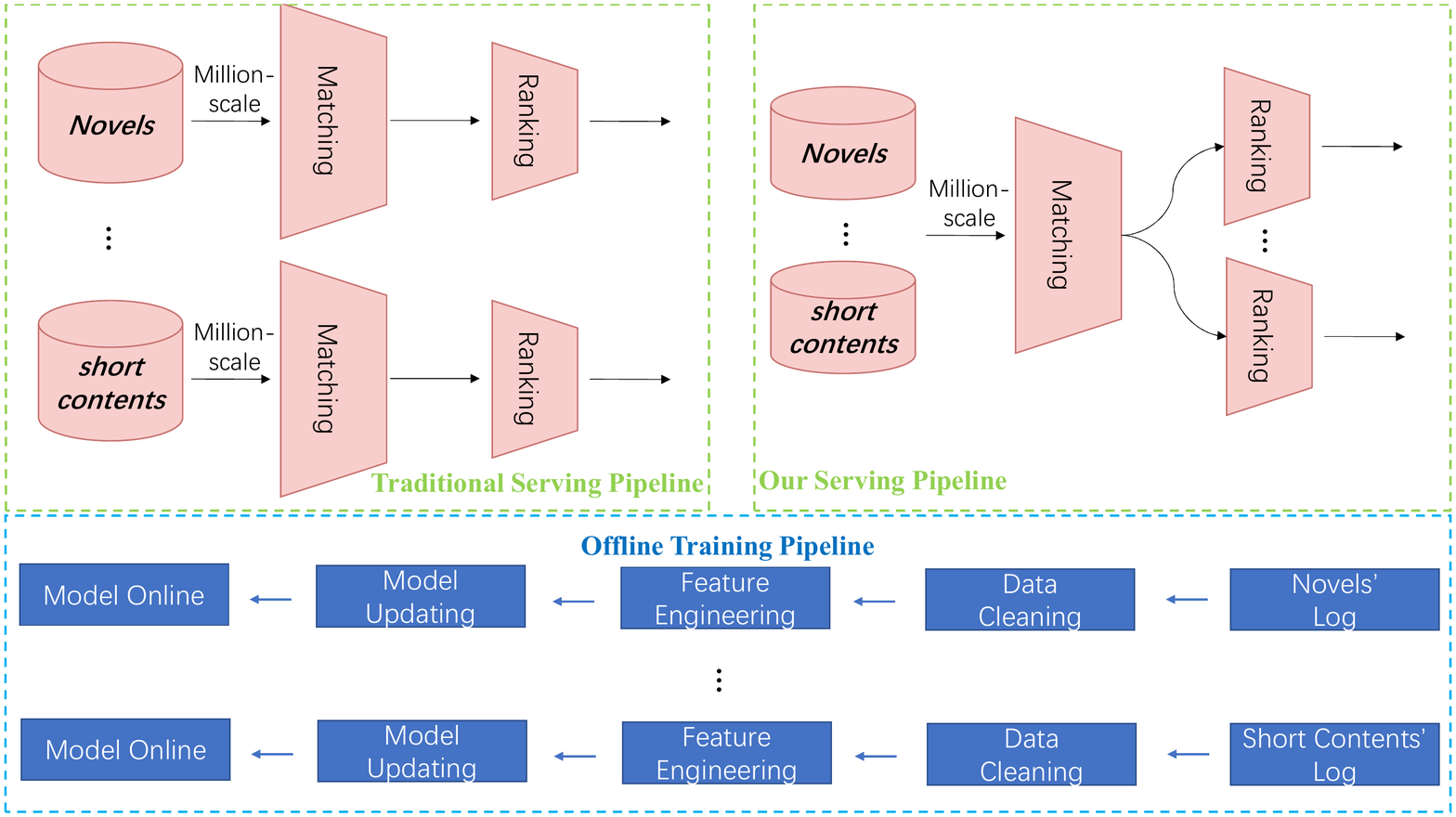}
\vspace*{-1.5em}
\caption{The online and offline pipelines of the industrial recommender  systems.}
\label{fig:pipeline}
\end{figure}

Personalized recommender systems have been widely employed in web applications, such as e-commerce services, news recommendations, and video recommendations ~\cite{covington2016deep, sun2021group, chen2022meta, gefan2022field}. The recommender systems improve user experience by filtering items in which users are interested. In the industrial scenario, scoring large-scale items effectively in real-time becomes  challenging, as the  system serves billion-scale users with billion-scale contents, e.g., QQ Kandian. There is a general practice that designs the whole system with a matching stage and a ranking stage, \fg{as shown in Figure \ref{fig:pipeline}}. The matching stage aims at retrieving hundreds or thousands of satisfying items from billions of candidates. Next, the ranking stage generates a meticulous ranking list based on the selected items. \fg{Both matching and ranking stages play critical roles in the entire recommendation pipeline. In this work, we target on the matching stage. }

Inspired by recent success of deep learning achieved in other domains \cite{mikolov2013distributed,he2016deep,vaswani2017attention,zhang2020microscope,zhang2022quickskill}, many research applies deep neural networks (DNNs) to personalized recommender systems~\cite{covington2016deep,chen2022meta,fan2022pppne}. In particular, the two-tower model (TTM) is one of the most popular DNN-based methods deployed in the real-world matching stage ~\cite{huang2013learning, yi2019sampling, huang2020embedding}. TTM learns two mappings for users and item features via two independent DNN towers, encoding representations of users and items in the same space. This is beneficial to the online serving module, as similar items can be efficiently retrieved with an approximate nearest neighbor (ANN) algorithm ~\cite{johnson2019billion}. However, though the TTM-based models are effective, there remain several challenges to be addressed. Namely,
\begin{itemize}
    \item \textbf{Information Sharing.} Modern information recommender applications usually contain various types of content, such as short news, novels, images, and videos. Since  distributions of features and labels vary from different content types, traditional industrial recommender systems serve different types with corresponding models. This limits the information sharing across different data sources. 
    \item \textbf{Contents Cold Start.} Since many of the contents are created frequently by users in online applications, the cold start problem is exacerbated. Current recommender  systems not only meet the cold start challenges for new content, but also new features and data types.
    \item \textbf{Feature Interaction.}  A large number of models utilize vanilla multilayer perceptrons (MLPs) to learn the high-order feature interactions. However, MLPs are inefficient in dealing with multiplicative high-order feature interactions.
    \item \textbf{Data Flow Asynchronism.} In real-world recommender  systems,   machine learning models are updated frequently, so as to capture the evolution of users' interests agilely. Nevertheless, different contents in the data flow are processed with various pipelines, which causes asynchronism in terms of data updates. The modern recommendation model should be capable of handling asynchronous data flows at both training and inference stages.
    
\end{itemize}

Though there exists some research attacking these issues (e.g., \cite{song2019autoint,wang2019improving,lin2021transfer,fan2021predicting}), only few of them address all problems simultaneously, and are deployed in real-world large-scale recommender systems.

To tackle the above challenges, we propose Multi-View Hybrid Attentive Networks (MV-HAN), for the matching stage in industrial recommender systems. The MV-HAN extends the TTM by transferring information between different content types via several hybrid neural networks. The proposed method shares parameters of the bottom structures between different data types, which enables information sharing effectively. As such, the minor data types and features are both well-trained. Our MV-HAN is implemented with several multi-head self-attentive neural networks with residual connections, which promotes feature interactions. This enables knowledge to be transferred efficiently among different content types, \fg{and allows to automatically process multiple types of data from asynchronous pipelines.}

To summarize, this paper makes the following contributions:
\begin{itemize}
    \item We proposed a generic modeling framework for the matching stage in the recommender systems by extending the two-tower model for transferring source information to the target type.
    
    \item We introduce an alternate training algorithm to optimize multiple objects concurrently, which enables the proposed MV-HAN to optimize easily between different datasets. 

    \item We test our proposed model in offline experiments with top-N recommendations. Experiments show that MV-HAN outperforms state-of-the-art baselines and achieves up to $4.64\%$ higher Hit Ratio (HR) score. 
    \item We deploy the MV-HAN in a real-world recommender system. The online A/B test shows that the proposed model obtains significant improvements in all metrics. These results demonstrate the efficiency of our design.
    
\end{itemize}

\section{Proposed Method}
\subsection{Problem Statement}
In this work, we focus on the matching stage in recommender  systems.
The issue in the matching stage is a typical Information Retrieval (IR) question, aiming at retrieving a set of content that users are interested in from massive content. We consider a user set $\mathbf{U} = [u_1,u_2,....,u_{N_u}]$, a source content set $\mathbf{O}^{s} = [o^{s}_1,o^{s}_2,....,o^{s}_{N^{s}_v}]$ and a target content set $\mathbf{O}^{t} = [o^{t}_1,o^{t}_2,....,o^{t}_{N^{t}_v}]$, where $N_u$, $N^{s}_v$  and $N^{t}_v$ denote the number of users, source contents and target contents respectively. 
The user-content interaction history can be defined as a matrix $\mathbf{Y} \in \mathbb{R}^{ N_u \times (N^{s}_v + N^{t}_v)} $, where $y_{ij} = 1 $ if the interaction between user $i$ and content $j$ is observed, and $y_{ij} = 0 $ otherwise. Our object is to learn a function $ \hat{y}_{ij} =  f (u_i, o_j \vert \mathbf{\Theta})$ to predict the score $\hat{y}_{ij}$ of interaction $y_{ij}$, where $\mathbf{\Theta}$ denotes model parameters. In this way, we can rank relevant content by prediction scores. 

\subsection{Multi-View Hybrid Neural Networks}
\begin{figure*}[!tp]
\centering
\includegraphics[width=14cm]{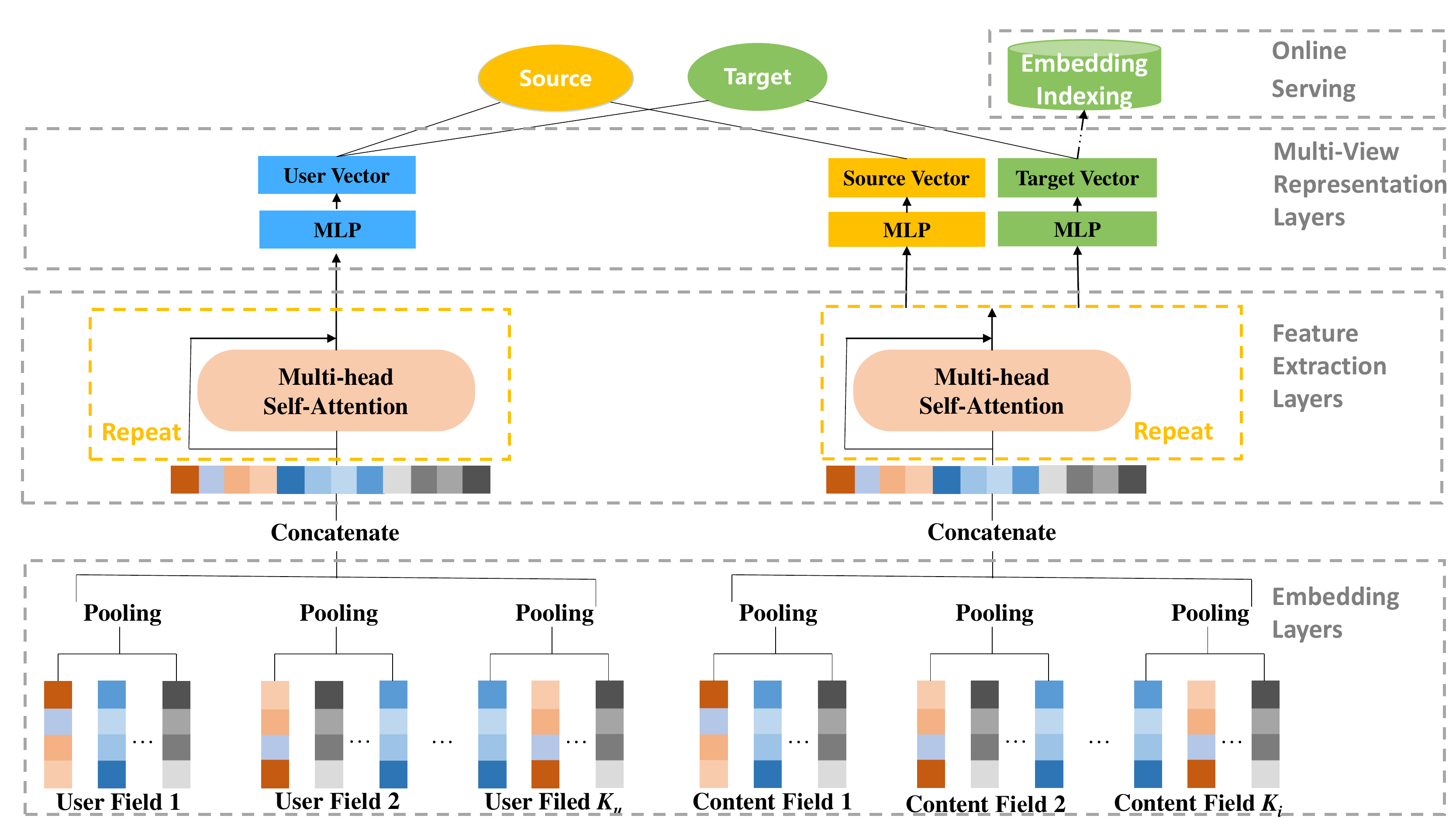}
\caption{An overview of the proposed MV-HAN.}
\label{fig:framework}
\end{figure*}

We show the overall architecture of the MV-HAN in Figure \ref{fig:framework}. The MV-HAN contains two model towers for users and contents respectively. This inherits from TTM to keep the embedding of users and contents independently and retrieves similar contents in the online serving stage efficiently. Each tower of the MV-HAN includes three major structures: embedding layers, feature extraction layers, and multi-view representation layers. The users' tower can learn better representations by sharing parameters even if the data of the target types is limited.
Different from users' towers, the content tower merely shares  parameters in  embedding layers and  feature extracting layers. This is to balance the learning process between different types of content to mitigate the cold start issue in the sparse types. 
In addition, feature extraction layers are implemented with multi-head self-attentive neural (MHSA) networks  with residual connections, which extract high-order latent feature interactions effectively ~\cite{vaswani2017attention, song2019autoint}. The multi-view representation layers are several independent MLPs, capturing the difference between source and target contents. 
Finally, the MV-HAN predicts the final scores by the users' and contents' representations. We formulate overall MV-HAN as follows:
\begin{equation}
\label{eq:embedding}
\begin{aligned}
z^u_i &= f_u(p_i) = MRL_u(FEL_u(p_i)), \\
z^s_j &= f_s(q^{s}_i)= MRL^{s}_o(FEL_o(q^{s}_j)), \\
z^t_k &= f_t(q^{t}_k)= MRL^{t}_o(FEL_o(q^{t}_k)). \\
\end{aligned}
\end{equation}
Here $z^u_i$ denotes the $i$-th user's representation, $z^s_j$, and $z^t_k$ denote the $j$-th, and $k$-th content's representations of source and target data. $MRL_u$, $MRL^{s}_o$, and $MRL^{t}_o$ denote the mappings of multi-view representation layers with users, source contents, and target contents, which are implemented by several MLPs. $FEL_u$ and $FEL_o$ denote the mappings of feature extraction layers, which are implemented by multiple blocks via MHSA networks with residual connections. The $p_i$, $q^{s}_i$, and $q^{t}_k$ are formulated as:
\begin{equation}
\label{eq:combine}
\begin{aligned}
p_i = Emb_u(u_i), \quad
q^{s}_j = Emb_o(o^{s}_j), \quad
q^{t}_k = Emb_o(o^{t}_k),
\end{aligned}
\end{equation}
where $Emb_u$ includes the embedding and concatenate function for users, and $Emb_o$ for contents.

We follow a common setting to predict the user-content interaction scores by the cosine function. Specifically:
\begin{equation}
\label{eq:finalRep}
\begin{aligned}
\hat{y}^{s}_{ij}  = \frac{z^{uT}_i z^s_j}{\Vert z^u_i \Vert \Vert z^s_j \Vert}, \quad
\hat{y}^{t}_{ik} = \frac{z^{uT}_i z^t_j}{\Vert z^u_i \Vert \Vert z^t_k \Vert },
\end{aligned}
\end{equation}
where $\hat{y}^{s}_{ij}$ and $\hat{y}^{t}_{ij}$ denote the prediction score of the source type and the target type.
\subsection{Optimization Objective}
The retrieval problem is essentially a classification problem, which estimates the probability distribution of $\hat{y}$ with a softmax function:
\begin{equation}
\label{eq:softmax}
P(y^{s}_{ij} \vert \hat{y}^{s}_{ij}) = \frac{\exp(\hat{y}^{s}_{ij})}{\sum^{N^{s}_i}_{j = 1} {\exp(\hat{y}^{s}_{ij}}) }, \quad
P(y^{t}_{ik} \vert \hat{y}^{t}_{ik}) = \frac{\exp(\hat{y}^{t}_{ik})}{\sum^{N^{t}_i}_{j = 1} {\exp(\hat{y}^{t}_{ik}})}.
\end{equation}
The loss for user $i$ can be formulated as:
\begin{equation}
\label{eq:loss}
\begin{aligned}
L^{s}_i = \sum^{N^{t}_i}_{j = 1}( y^{s}_{ij}\log{P(y^{s}_{ij} \vert \hat{y}^{s}_{ij}})),\quad
L^{t}_i = \sum^{N^{t}_i}_{k = 1}( y^{t}_{ik}\log{P(y^{t}_{ik} \vert \hat{y}^{t}_{ik}})).
\end{aligned}
\end{equation}
Since $N_j$ is usually huge in industrial scenarios, it is  time-consuming to include all contents in computing Eq. \ref{eq:loss}. To address this problem, for each user-content interaction, we randomly sample $r$ negative cases in $y_{ij} = 0$ for user $i$. Then we use Stochastic Gradient Decent (SGD) and its variants to optimize multiple objectives. We employ an alternative way to train the proposed method. Specifically, we put pairs with the same type in a batch, e.g., the source pair $(u_i,o^s_j ,y^s_{ij})$, and update model parameters with this batch. Next, we train the model with data of another type to update model parameters.

\section{Experiment}
We conduct both online and offline experiments to evaluate the performance of the proposed MV-HAN, \fg{to answer the following research questions: }
\begin{itemize}
    \item \textbf{RQ1:} Can our proposed MV-HAN outperform the state-of-the-art methods at the matching stage?
    \item \textbf{RQ2:} How do the key designed structures affect the performance of our proposed MV-HAN?
    \item \textbf{RQ3:} How does MV-HAN handle the online serving and perform for a real-world application?
\end{itemize}

To answer RQ1, we conduct offline experiments on several industrial datasets to compare the MV-HAN with baselines.

To answer RQ2, we conduct ablation studies to evaluate our purpose-built designs.

To answer RQ3, we deploy MV-HAN in our real-world application and conduct an online A/B test to evaluate the performance of the proposed model.

\subsection{Offline Experiment}
\subsubsection{Datasets}
We collect large-scale industrial datasets from QQ Kandian with different types. We use the data in the first nine days for the train sets and the rest are used for the test set. The datasets include different data types collected in different sources, where we show their statistics of datasets in Table \ref{tab:dataset}.  Specifically:
\begin{itemize}

\item \textbf{Articles} is a mature type in our platform, including texts and images. We choose it as the source dataset since it contains rich information about users.

\item \textbf{Novels} is a new type in QQ Kandian. We collect user behaviors from this type as the target dataset to evaluate the performance of methods.

\item \textbf{Short Contents (SC)} is a relatively sparse type since short contents usually are released and taken down fast.  The data of short contents are selected as a different target source. %
\end{itemize}

\begin{table}[t]
\caption{Statistics of the benchmark dataset. 
}
\centering
\begin{tabular}{l c c c ccc }
\hline
Dataset& Max Features Size &  \#Category & \#Samples \\
\hline
\hline
Articles& $ \sim  10^8$& 13 & $ \sim 2 \times 10^7$\\
Novel& $ \sim  10^7$&  13 & $ \sim 3 \times 10^6$\\
SC& $ \sim 10^7$ &  13 & $ \sim 9 \times 10^6$ \\
\hline
\end{tabular}
\label{tab:dataset}
\end{table}

\subsubsection{Baselines Methods.}
We compare the proposed method with several existing recommendation methods used in the matching stage, including logistics regression (LR) ~\cite{mcmahan2013ad}, YoutubeDNN ~\cite{covington2016deep}, Two-Tower based Model (TTM) ~\cite{yi2019sampling}, Cross-domain Content-boosted Collaborative Filtering neural NETwork (CCCFNet) ~\cite{lian2017cccfnet} and  Multi-View Deep Neural Network (MV-DNN) ~\cite{elkahky2015multi}. We train TTM on both source and target datasets to evaluate the performance of knowledge transfer. We mark this method as TTM\_all. 

\subsubsection{Metrics}
We employ two widely used metrics, \emph{Area Under the ROC Curve} (AUC) and \emph{Hit Ratio (HR@50)} \cite{xie2020internal,lin2021transfer} to evaluate the performance of the MV-HAN. The AUC is used for evaluating the ranking performance of contents, and the HR is used for testing \emph{whether good contents are retrieved}. \fg{Note that we use the relative AUC improvement \cite{zhou2018deep,gefan2022field} for the evaluation in this paper.} 

\begin{table}[t]
\caption{Performance of all models considered in this study. The best results are boldface. The underlined values are the best results of baseline methods. $RelaImpr$ shows the relative improvement between MV-HAN with underlined values.  
}
\centering
\begin{tabular}{l c c c c }
\hline
&\multicolumn{2}{c}{SC}&\multicolumn{2}{c}{Novels}\\
\hline
&AUC&HR&AUC&HR\\
\hline
\hline
LR&0.6644 &0.1940 &0.7154 &0.1345 \\
YoutubeDNN&0.6681 &0.2385 &0.7169 &0.1624 \\
TTM&0.6699 &0.2233 &\underline{0.7232} &\underline{0.1653} \\
TTM\_all&0.6791 &0.2666 &0.7132 &0.1518 \\
CCCFNet&0.6075 &0.1383 &0.6743 &0.1043 \\
MV-DNN&\underline{0.6988} &\underline{0.2821} &0.7225 &0.1575 \\
\hline
MV-HAN&\textbf{0.7076} &\textbf{0.2952} &\textbf{0.7269} &\textbf{0.1718} \\
$RelaImpr$&4.43\% &4.64\% &1.66\% &3.93\% \\

\hline
\end{tabular}
\label{tab:result}
\end{table}

\subsubsection{Result Analysis}
We show the results in Table \ref{tab:result}. Observe that our proposed MV-HAN consistently outperforms all baseline methods on both datasets. Specifically, MV-HAN achieves up to 93.12\% higher relative AUC and 113.45\% higher relative HR. This verifies the strong ability of interested content retrieval for our model.
In addition, though TTM\_{all} and MV-DNN can utilize the source information \fg{ to mitigate the cold start problem}, the vanilla TTM outperforms those methods in the Novel dataset. This demonstrates that simply utilizing source data may dilute the knowledge of the target type. MV-HAN also achieves the best results in this dataset, which proves that the proposed method is more effective for transferring the latent information between different data types.

\subsubsection{Ablation Studies}
There are three key components in the proposed MV-HAH. We respectively conduct ablation studies to evaluate each of their effectiveness. Specifically, MV-HAN w/o SE is a variant of the MV-HAN, which uses exclusive embedding layers. MV-HAN w/o FE removes the shared feature extracting layers from the MV-HAN, while it employs independent networks for each type in the content tower. The MV-HAN$_{MLP}$ replaces multi-head self-attentive neural networks with MLPs for feature extraction. We show the results in Table \ref{tab:ablation}. Observe that the MV-HAN achieves the best performance in both datasets, which verifies the effectiveness of the purpose-built design.

Specifically, compared to the MV-HAN w/o SE, the MV-HAN obtains substantial relative improvements in SC over the Novel dataset. This is because the short contents are updated more agilely than in novels. Thus, the model becomes underfitting over this data type. In contrast, compared to the MV-HAN w/o FE, MV-HAN achieves greater relative improvements in Novel than in the SC dataset. As the novel type includes the smallest number of training data, the corresponding tower may be trained insufficiently. In addition, compared to MV-HAN$_{MLP}$, the MV-HAN also achieves 1.07\% $\sim$ 2.08\%  relative improvements in the two datasets. This demonstrates that the multi-head self-attentive neural networks have better feature extraction ability over MLPs.

\begin{table}[t]
\caption{Results of the ablation studies of our MV-HAN. $RelaImpr$ shows improvement compared with other models.}
\centering
\begin{tabular}{l c c c c }
\hline
&\multicolumn{2}{c}{SC}&\multicolumn{2}{c}{Novels}\\
\hline
&AUC&HR&AUC&HR\\
\hline
MV-HAN&0.7076&0.2952&0.7269&0.1718\\
\hline
MV-HAN w/o SE&0.7061&0.2915&0.7260&0.1697\\
$RelaImpr$&0.73\%&1.27\%&0.40\%&1.24\%\\
\hline
MV-HAN w/o FE&0.7049&0.2909&0.7229&0.1656\\
$RelaImpr$&1.32\%&1.48\%&1.79\%&3.74\%\\
\hline
MV-HAN$_{MLP}$&0.7035&0.2898&0.7245&0.1683\\
$RelaImpr$&2.01\%&1.86\%&1.07\%&2.08\%\\
\hline
\end{tabular}
\vspace*{-0.5em}
\label{tab:ablation}
\end{table}

\subsection{Online A/B Test}
In order to evaluate the performance of the proposed method in the real-world application, we deploy MV-HAN in the \emph{QQ Kandian short content recommendation} and conducted live experiments to compare MV-HAN with a two-tower based method only trained by the short content data. The online evaluation metrics employed for evaluation are \emph{Click-Through-Rate (CTR)}, the number of the \emph{Daily Active User (DAU)}, the number of the \emph{clicks} and the \emph{Duration} users used. Table \ref{tab:abtest} shows the relative changes after employing our method. Observe that the proposed MV-HAN obtains significant improvements in all metrics. Specifically, the MV-HAN achieves 6.95\% and 10.12\% relative improvement in terms of \emph{CTR} and \emph{\#Clicks}, respectively. This is because MV-HAN can transfer information from other types to the short content to retrieve contents in which users are more interested. \emph{DAU} and \emph{Duration} are regarded as key metrics for respectively evaluating the short-term and long-term competitiveness of a platform, especially for user-generated content (UGC) applications. The MV-HAN gains relative improvement of 5.09\% and 9.69\%  in \emph{DAU} and \emph{Duration} respectively. It shows that the MV-HAN is beneficial in keeping users for both the short term and the long term.

\begin{table}[tbp]
\caption{Relative changes for all metrics after employing the MV-HAN in online A/B Test.}
\centering
\begin{tabular}{l l| l l}
\hline
Metric&Change&Metric&Change\\
\hline
\hline
\#DAU&$+5.09\%$ &Duration& $+9.69\%$ \\
\#Clicks & $+10.12\%$ & CTR& $+6.95\%$\\
\hline
\end{tabular}
\vspace*{-1.5em}
\label{tab:abtest}
\end{table}

\section{Conclusion}
In this paper, we propose a novel model called MV-HAN for the matching stage in recommender systems. We design a hybrid neural structure configured with different models, including MLPs and multi-head self-attentive neural networks. The proposed method transfers the knowledge from the source types to the target types, which helps better representation learning for users and contents. Moreover, the MV-HAN shares parameters of the bottom networks to mitigate the cold start on the spare types. Offline experiment results on industrial datasets show that the proposed method outperforms different baselines, i.e., achieving up to 4.43\% and 4.64\% higher AUC and HR than the best results of baseline methods on the SC dataset. 
Online experiment results on real-world recommender  systems show that the MV-HAN significantly improves the recommendation performance compared with baseline methods in all metrics. \fg{It verifies that the MV-HAN is able to handle multi-source asynchronous dataflows and extract information from different content types in real-world applications. }

\section*{Acknowledgements }
We thank all anonymous reviewers for their hardworking  and suggestions.
This work was supported in part by the National Natural Science Foundation of China under Grant No. 62102265, by the Open Research Fund from Guangdong Laboratory of Artificial Intelligence and Digital Economy (SZ) under Grant No. GML-KF-22-29, by the Natural Science Foundation of Guangdong Province of China under Grant No. 2022A1515011474.





\bibliographystyle{ACM-Reference-Format}
\bibliography{sample-base}


\end{document}